\def\b{\bibitem}
\def\be{\begin{equation}}
\def\ee{\end{equation}}
\def\bea{\begin{eqnarray}}
\def\eea{\end{eqnarray}}
\def\bml{\begin{mathletters}}
\def\eml{\end{mathletters}}
\documentstyle[aps,prb,eqsecnum,psfig,epsf,floats]{revtex}
\draft
\begin{document}
\def\SNG{{\em Physical Review Style and Notation Guide}}
\def\LUG {{\em \LaTeX{} User's Guide \& Reference Manual}}
\def\btt#1{{\tt$\backslash$\string#1}}%
\def\REVTeX{REV\TeX}
\def\AmS{{\protect\the\textfont2
        A\kern-.1667em\lower.5ex\hbox{M}\kern-.125emS}}
\def\AmSLaTeX{\AmS-\LaTeX}
\def\BibTeX{\rm B{\sc ib}\TeX}
\twocolumn[\hsize\textwidth\columnwidth\hsize\csname@twocolumnfalse%
\endcsname
\title{Comment on ``Specific heat of a Fermi system near ferromagnetic quantum
       phase transition'' by Grosu, Bodea, and Crisan (cond-mat/0101392)}
\author{D. Belitz}
\address{Department of Physics and Materials Science Institute\\
         University of Oregon,
         Eugene, OR 97403}
\author{T.R. Kirkpatrick}
\address{Institute for Physical Science and Technology, and Department of 
         Physics\\
         University of Maryland,
         College Park, MD 20742}
\date{\today}
\maketitle
\begin{abstract}
\end{abstract}
]
In a recent paper,\cite{Grosuetal} Grosu et al. have considered the
thermodynamic behavior of an itinerant ferromagnet at the quantum
critical point. Specifically, they find that the leading critical
temperature dependence of the specific heat
coefficient in three-dimensional clean systems is logarithmic,
rendering the critical system a non-Fermi liquid.
We would like to point out that this result is not new, contrary to
the claim made in the paper. Rather, it is identical with the result
obtained by Millis\cite{Millis} in 1993 (the authors' Reference [7]).
Furthermore, we note that it has been shown in Ref. \onlinecite{us_1st_order}
that the ferromagnetic quantum phase transition in a clean system is
generically of first order, so the authors' calculation is not applicable 
to this problem.

Apart from this failure to reference prior results, the paper's
discussion section contains the following unsupported and incorrect claims.

(1) The integrating out of the fermions in Refs. \onlinecite{us_dirty,us_clean}
(the second reference in the authors' Ref. [12], and their Ref. [14],
respectively) is claimed to neglect the coupling between spin fluctuations 
and fermionic particle-hole excitations. This is not correct. The integration
over the fermionic degrees of freedom is an exact prodecure, and the
physical effects in question are contained in the vertex functions of the
resulting Landau-Ginzburg-Wilson theory. In fact, it is precisely the coupling
to particle-hole excitations that produces the nonanalyticity in the
paramagnon propagator which the authors object to, see point (2) below. 
Conversely, this means that the authors of Ref.\ \onlinecite{Grosuetal}
fail to take into account any nontrivial correlations from the fermionic
degrees of freedom, since they take Hertz's action for granted. These
correlations are responsible, among other effects, for the first order
nature of the transition.\cite{us_1st_order}
The remaining papers quoted in the
authors' Refs. [11-14] did not integrate out the fermions, contrary to
the authors' claim, and the point is therefore moot.

(2) The authors claim, without explanation, that the nonanalyticity in
the paramagnon propagator in Refs. \onlinecite{us_dirty,us_clean} violates
spin-density conservation. This is not correct. In fact, considering the
paramagnon propagator
$$
P({\bf k},\Omega_n) = \frac{1}{t + f({\bf k}) 
   + \vert\Omega_n\vert/\vert{\bf k}\vert^n}
   \quad,
$$
spin-density conservation is guaranteed by the term 
$\vert\Omega_n\vert/\vert{\bf k}\vert^n$ ($n=1,2$ for clean, disordered sytems,
respectively), irrespective of whether or not the function $f({\bf k})$ is
analytic at ${\bf k}=0$.

(3) Contrary to the authors' claim, the value of the dynamical exponent $z$
is not an open problem; it has been discussed at length in their 
Refs. [6,7,12,13].

\acknowledgments
This work was
supported by the NSF under grant Nos. DMR-98-70597 and DMR-99-75259.

\end{document}